# Surface-enhanced Raman scattering and density functional theory study of selected-lanthanide-citrate complexes (lanthanide: La, Ce, Pr, Nd, Sm, Eu, and Gd)

Hao Jin[1], Tamitake Itoh[2] and Yuko S. Yamamoto[1]*

[1] School of Materials Science, Japan Advanced Institute of Science and Technology (JAIST), Nomi, Ishikawa 923-1292, Japan

[2] Nano-Bioanalysis Research Group, Health Research Institute, National Institute of Advanced Industrial Science and Technology (AIST), Takamatsu, Kagawa 761-0395, Japan

*Corresponding author: yamayulab@gmail.com

## Abstract

In this study, we combined the surface-enhanced Raman scattering (SERS) with density functional theory (DFT) calculations to investigate the SERS spectra of lanthanide (Ln)–citrate complexes (Ln = La, Ce, Pr, Nd, Sm, Eu, and Gd) under 488, 532, and 660 nm laser excitations. Detailed vibrational analysis and peak assignments were performed based on SERS spectra simulated using an optimized DFT setting, in which small-core effective core potentials (ECPs) in the def2-tzvpd basis set were replaced by large-core ECPs. Characteristic SERS peaks appeared at 1065, 1315, and 1485 $cm^{-1}$ were assigned to the $\gamma(CH_2) + \nu(C–O\cdots Ln)$, $\nu_{sym}(COO^-) + \gamma(CH_2)$, and $\nu_{asym}(COO^-) + \gamma(CH_2)$ vibrational bands, respectively. SERS intensity ratios were obtained by normalizing the peak intensity I near 1065 or 1485 $cm^{-1}$ to that near 1315 $cm^{-1}$. $I_{1065}/I_{1315}$ depended solely on the type of $Ln^{3+}$ ion and was independent of the excitation wavelength. In contrast, $I_{1485}/I_{1315}$ increased with decreasing excitation wavelength, indicating additional enhancement by charge-transfer. Additionally, as the number of unpaired 4f electrons increased, $Ln^{3+}$ in the coordination region attracted oxygen negative charges more strongly, reducing the electric dipole moment of the C–O bond and altering its symmetry.

## Introduction

Lanthanides (Ln), comprising 15 elements with atomic numbers from 57 (La) to 71 (Lu), are crucial rare earth elements[1]. As the atomic number increases, electrons gradually fill the 4f orbitals of Ln[1]. Owing to the electronic configuration of $Ln^{3+}$ ions, which can be expressed as [Xe]$4f^n$ (n=0−14). Lanthanides exhibit similar chemical properties while possessing abundant energy levels and high-spin states. This unique electronic configuration results in exceptional optical and magnetic properties.[1-7] , which have led to the widespread use of $Ln^{3+}$ ions and their small molecular complexes in scientific research. For instance, they not only play an important role in functional materials developments, such as the synthesis of luminescent materials[2,3], biomedical probes[3,4], and magnetic resonance imaging (MRI) contrast agents[5], but are also used in quantum computing[6] and spin detection[7]. Furthermore, they are utilized to explore special enhancement phenomena, such as magneto-chiral dichroism enhancement[8] in circularly polarized luminescence[9]. Additionally, lanthanides exhibit a significant characteristic wherein increasing the atomic number enhances the attraction between the nuclear charge and outer electrons due to the weak shielding effect of f-electrons, causing a gradual decrease in the ionic radius of $Ln^{3+}$, known as lanthanide contraction[1,10,11].

High-performance measurement methods are required to measure the characteristic signals of $Ln^{3+}$ ions effectively. However, measuring low-concentration ($<10^{-5}$M) samples and rapidly obtaining structural information about Ln-molecular complexes using conventional techniques such as nuclear magnetic resonance (NMR)[12], fluorescence[2-4], and electron spin resonance (ESR) spectroscopy[13], remain challenging. Therefore, developing more sensitive and efficient techniques for measuring the characteristic signals of $Ln^{3+}$ ions, particularly in low-concentration($<10^{-5}$M) samples, is an important research focus. Raman spectroscopy is a non-destructive technique that provides molecular fingerprints[14] and vibrational information[15] on Ln-molecular complexes. When a molecule is adsorbed on a noble metal surface, its Raman intensity is enhanced by electromagnetic or chemical effects, such as charge-transfer (CT).[16] This phenomenon known as surface-enhanced Raman scattering(SERS)[16-18], enables the detection of low concentrations ($<10^{-5}$M) and even single molecules [18]. Thus, SERS is a valuable method for measuring the characteristic signals of $Ln^{3+}$ ions and can be integrated with other standard techniques for multimodal measurement.

However, due to the similar chemical properties of $Ln^{3+}$ ions, their molecular complexes also exhibit similar chemical structures, rendering their vibrational spectra difficult to distinguish[19]. Moreover, because the 4f orbitals are shielded by the 5s and 5p orbitals, they provide a wealth of energy levels while not directly participating in bonding[1]. Consequently, certain $Ln^{3+}$ ions with multiple energy levels[20], such as $Pr^{3+}$, $Nd^{3+}$, and $Eu^{3+}$, may exhibit fluorescence interference in Raman measurements [21,22] or induce resonance Raman effects due to energy absorption [23], further complicating the analysis of Ln-molecular complexes. Thus, the application of SERS to Ln-molecular complexes is limited to some extent, primarily focusing on individual $Ln^{3+}$ ions. Examples include, investigating the effect of $Nd^{3+}$ ions on semiconductor SERS substrates [24], exploring SERS enhancement mechanisms in Gd-complexes [25], and developing MRI-SERS multimodal Gd-fluorescent probes [26]. Our previous research successfully classified $La^{3+}$ and $Gd^{3+}$ ions, which have significant spin differences, using SERS technique[19]. However, the feasibility of using SERS to distinguish other $Ln^{3+}$ ions, particularly those with smaller differences in electronic configuration, such as $Pr^{3+}$ and $Nd^{3+}$, remains unverified. Additionally, because of the complex 4f electronic configuration of $Ln^{3+}$ ions and pronounced relativistic effects[27], conventional density functional theory (DFT) methods face significant challenges in simulating the SERS spectra of Ln-molecular complexes, although we partly succeeded in simulating the SERS spectra of La- and Gd-citrate complexes[19]. Therefore, simpler and more efficient SERS simulation methods must be developed to assist in the analysis of experimental SERS spectra and advance the study of Ln-molecular complexes.

In this study, we investigated the relationships between the characteristic peaks in the SERS spectra of Ln-citrate complexes using a combined experimental and computational approach. Ln with 0-7 (excluding 4) unpaired 4f electrons, representing La, Ce, Pr, Nd, Sm, Eu and Gd, respectively, were selected for this study. Pm, with 4 unpaired 4f electrons, was excluded from the study because of its radioactivity and instability. The SERS spectra of the Ln-citrate complexes were obtained using citrate-capped silver nanoparticle (citrate@AgNP) colloids containing the corresponding $Ln^{3+}$ ions added. Using DFT-simulated SERS spectra for peak assignment, we conducted a detailed analysis of the differences in the SERS spectra and possible mechanisms underlying these differences. These results provide valuable insights for developing new methods for SERS measurement and simulation of Ln-molecular complexes, as well as for exploring the SERS and coordination chemistry of $Ln^{3+}$ ions.

## Experiments

**Chemicals.** Six different lanthanide oxides were used to prepare $Ln^{3+}$ solutions. $La_2O_3$, $Pr_6O_{11}$, $Nd_2O_3$, and $Gd_2O_3$ were sourced from Fujifilm Wako Pure Chemical Corporation (Japan), while $Sm_2O_3$ and $Eu_2O_3$ were obtained from Kanto Chemical Co., Inc (Japan). All lanthanide oxides had a purity of > 99%. Ultrapure water from a Direct-Q® UV 3 system (Millipore, USA) was used as the solvent. $Ln(NO_3)_3$ (Ln: La, Pr, Nd, Sm, Eu, and Gd) solutions with a concentration of 0.2 M were prepared by dissolving the corresponding lanthanide oxides in 1 M $HNO_3$ (Fujifilm Wako Pure Chemical Corporation, Japan) under heating. These mother solutions were then diluted with ultrapure water to obtain $2\times10^{-3}$ M $Ln(NO_3)_3$ solutions. A 0.2 M $Ce(NO_3)_3$ solution was prepared by dissolving $Ce(NO_3)_3$ crystals (99%, Fujifilm Wako Pure Chemical Corporation, Japan) in ultrapure water. Citrate@AgNPs, colloidal AgNPs dispersed in water and stabilized by surface-bound citrate molecules, were prepared using a modified Lee & Meisel method as described in our previous study[19].

**Characterization.** Samples for SERS measurements were prepared by adding 50 µL of the $2\times10^{-3}$ M $Ln(NO_3)_3$ solution to 1 mL of citrate@AgNPs, obtaining a concentration of $1\times10^{-4}$ M for each Ln-citrate@AgNPs sample. The samples were stored at room temperature for 12 hours prior to spectroscopic measurements.

For the ultraviolet-visible (UV-Vis) measurements, each SERS sample was diluted 10-fold with ultrapure water, and the measurements were conducted using a UV-Vis-NIR spectrometer (V-770, JASCO, Japan) with a 1 cm path-length polystyrene cuvette. Ultrapure water was used as the blank liquid.

SERS measurements were conducted using clean soda glass capillaries (1.1 ×75 mm,

DWK Life Sciences, USA) to hold the sample solutions, which were allowed to stabilize for approximately 8-12 hours before measurement. SERS spectra were then measured using a Raman microscope (T64000, Horiba Scientific, Japan) equipped with a 90X objective using 488, 532 and 660 nm laser at power levels of 50, 50, and 100 mW, respectively. The corresponding energy powers at the sampling points were 101, 85, and 110 mW $\mu m^{-2}$, respectively. To avoid sample damage from high-power laser exposure and ensure repeatability, the exposure time for the SERS measurements was set to 30 seconds with 2 accumulations. Raman shift correction was performed using indene.

Spectral analysis was carried out using OriginPro 2022 (OriginLab Corporation, USA). Baseline removal from the raw spectra was achieved using the interpolation mode, followed by noise reduction using the Savitzky-Golay method applied twice with an 11-point window and polynomial order of 1.

**Computational methods** The simulated SERS spectra of the Ln-citrate complexes on a silver surface were calculated using a simplified model based on our previous work[19]. The Ln-citrate complexes were modeled as $LnCit^-$ using the trisodium citrate structure, which was obtained from the Cambridge Crystallographic Data Centre (CCDC, http://www.ccdc.cam.ac.uk, deposition number: 1478188). To represent the silver surface, a cluster of 11 Ag atoms, approximating the size of the $LnCit^-$ complex, was extracted from a bulk Ag crystal along the (111) plane[28].

Geometry optimization and vibrational spectra calculations were performed using DFT with accurate Hessian matrices in the Gaussian 16 software(Version C.01, Gaussian Inc. USA)[29]. Vibrational and structural analysis of the optimized systems were performed using the GaussView software (Version 6, Gaussian Inc. USA)[30]. The commonly used B3LYP functional is not suitable for transition metals[31,32] and lanthanides[33,34]; the PBE0 functional[35] has been reported to provide better results for Ln-molecular complexes[36] and small molecules[37]. Thus, we employed the PBE0 functional with DFT-D3(BJ) dispersion correction[38] to accurately describe the Ln elements. The solvent environment was modeled as water using the SMD model[39]to replicate the experimental conditions. Considering the significant influence of basis sets on the geometry optimization of trisodium citrate and the Ln-citrate-$Ag_{11}$ complexes, the aug-cc-pvdz basis set[40] was used for C, H, O, and Na, and def2-svpd[41] and def2-tzvpd basis sets[42] for Ag and Ln, respectively. Due to convergence difficulties with some Ln-citrate complexes, large-core relativistic effective core potentials (ECPs)[43] were consistently used for all $Ln^{3+}$ ions. A large-core ECP replaces all electrons of the ion except for the 10 valence electrons in the 6s, 5s, and 5p orbitals. Thus, based on the number of inner electrons of the $Ln^{3+}$ ion replaced by the ECP, the scalar relativistic large-core ECPs MWB47, MWB48, MWB49, MWB51, MWB52, MWB53, and MWB54 were employed for $Ce^{3+}$, $Pr^{3+}$, $Nd^{3+}$, $Sm^{3+}$, $Eu^{3+}$, and $Gd^{3+}$, respectively, to replace the corresponding small-core ECPs in the def2-tzvpd basis set. The simulated Raman spectra of Ln-citrate-$Ag_{11}$ (simulated SERS spectra of Ln-citrate) were obtained by converting, the Raman scattering activity of each vibrational mode to the Raman intensity I using the following equation[44].

$$I_i = \frac{C(v_0 - v_i)^4 S_i}{v_i B_i}; B_i = 1 - \exp\left(-\frac{h v_i C}{k_B T}\right) \qquad (1)$$

where *i* is the vibrational mode, C is the normalization factor that can be arbitrarily chosen, *v* is the vibrational frequency, $v_0$ is the frequency of incident light. *S* is Raman activity calculated by Gaussian 16. h, c, $k_B$, and T are Planck constant, light speed, Boltzmann constant, and temperature, respectively. In the present work, the values of $v_0$ and T were chosen to be 532 nm and 298.15K, respectively, to match the experimental conditions.

The Raman shifts in the simulated SERS spectra were scaled by scaling factors of 0.956 and 0.971 for PBE0-D3(BJ)/aug-cc-pvdz and PBE0-D3(BJ)/def2-tzvpd, respectively, as obtained from the literature[45,46]. The Raman shifts in the region of 1000−1700 $cm^{-1}$ were scaled by a factor of 0.956, and then those in the region of 1000−1250 $cm^{-1}$ (after the first scaling) were scaled by a factor of 0.971 to account for the effects of the $Ln^{3+}$ ions. The full width at half maximum of all simulated SERS spectra was set to 25 $cm^{-1}$ to approximate the experimental spectra.

Additionally, energy calculations of the Ln-citrate-$Ag_{11}$ complexes(Ln: La and Gd) were performed at the PBE0-D3(BJ)/def2-tzvpp[47] level the small- and large-core ECPs to determine the effect of ECP-size on energy. The all-electron basis, x2c-tzvppall[48] were also used for the those energy calculations with DKH4 Hamiltonian[49]. All DFT calculation results were analyzed using the Multiwfn software[50,51] (3.8-dev version, Beijing Kein Research Center for Natural Sciences, China).

## Results and discussion

### SERS measurements of Ln-citrate complexes (Ln: La, Ce, Pr, Nd, Sm, Eu, and Gd).

Figure 1a shows a schematic diagram of the research methods used in this study. When individual $Ln^{3+}$ ions are added to citrate-capped AgNP colloids, they coordinate to the citrate molecules on the AgNP surface to form Ln-citrate complexes (denoted as Ln-citrate@AgNP, where AgNP serves as the core and Ln-citrate as the shell). Simultaneously, the addition of individual $Ln^{3+}$ ions induces the aggregation of AgNPs. In these aggregates, a strong electromagnetic field is generated by the localized surface plasmon resonance (LSPR) effect when the distance between the AgNPs becomes sufficiently small. The regions with such intense electromagnetic fields are referred to as "hotspots."[16-18] When Ln-citrate complexes are located within these hotspots, their Raman signals are significantly enhanced. To investigate the difference between the SERS of the Ln-citrate complexes, we performed SERS measurement, UV-Vis measurements, and SERS simulations The SERS spectra were simulated by computer modeling of simplified Ln-citrate@AgNP structures (Ln-citrate-$Ag_{11}$ complexes). As an example, the simulated and experimental SERS spectra of Gd-citrate is shown in Figure 1b. The close resemblance between the two spectra enabled the use of the simulated spectrum for vibrational mode analysis and peak assignment. The UV-Vis extinction spectra were also measured to check both the absorption and scattering of light by the AgNPs. Figure 1c shows the UV-Vis extinction spectra of the SERS samples diluted 10-fold with ultrapure water. The strong extinction at approximately 400 nm is primarily due to LSPR, where free electrons collectively oscillate in response to light[16]. Thus, the gradual decrease in extinction can be explained by the aggregation of AgNPs as $Ln^{3+}$ ions are added. Table 1 summarizes the absorption[52], and emission wavelengths[21] of the different $Ln^{3+}$ ions, along with their electronic configuration, ionic radii, and spin[1],

We combined experimental measurements with DFT-based peak assignments to analyze the SERS spectra of Ln-citrate complexes. Figures 2a–2c show the experimental

SERS spectra collected at 488, 532, and 660 nm, respectively, over the 450–2000 $cm^{-1}$ range, with most signals originating from citrate. Although Nd-citrate and Pr-citrate might exhibit pre-resonance Raman effects[23] at 488 nm and 532 nm, respectively (Table 1), their spectra did not changes, likely because the resonance in the coordination region was not pronounced. Similarly, the absence of fluorescence for $Sm^{3+}$ and $Eu^{3+}$ can be attributed to quenching[53,54] on the AgNP surface. Consistent with our previous study[19], the overall similarity in the spectral profiles of the Ln-citrate complexes is attributed to their similar structures. Detailed peak assignments were performed with the aid of DFT simulations (Figure 3 and Table 2). In the 450-1000 $cm^{-1}$ region (Figure 2), the small peaks around 620 and 700 $cm^{-1}$ were assigned to different $\delta(COO^{-})$, peaks around 740,820 and 945 $cm^{-1}$ were assigned to $\delta(COO^{-})$ + $\gamma(COO^{-})$, ν(CCCC–O) and $\nu(C\text{–}COO^{-})$ + δ(CH2), respectively. Under 488 and 532 nm excitation, the relative intensity differences among these peaks were not significant; however, under 660 nm excitation, the peak near 740 $cm^{-1}$ exhibited a significantly higher relative intensity than the other peaks in this region. In the 1000–1700 $cm^{-1}$ region, the large peaks at approximately 1065, 1315, and 1485 $cm^{-1}$ were ascribed to $\gamma(CH_2)$ + ν(C-O···Ln), $\nu_{sym}(COO^{-})$ + $\delta(CH_2)$, and $\nu_{asym}(COO^{-})$ + $\gamma(CH_2)$, respectively. The peak positions slightly varied depending on both the $Ln^{3+}$ species and excitation wavelength. This can be attributed to a secondary enhancement effect arising from electromagnetic (plasmon resonance variations) and chemical (CT related bond orientation) contributions[16,17,19]. Complementarily, Figure 3 presents the experimental SERS spectra of the Ln-citrate complexes under 532 nm excitation, alongside their simulated spectra and the optimized structures of Ln-citrate–$Ag_{11}$ clusters. The simulated spectra accurately reproduced the principal peaks observed experimentally beyond 1000 $cm^{-1}$, despite some discrepancies in the other regions. These differences can be attributed to several factors: (i) Experimental SERS signals arise from multiple enhancement mechanisms and are influenced by the laser frequency, whereas simulated spectra are computed under a static external field (zero frequency)[55] representing the ground-state CT effect[16,17]; (ii) The simulation considers only a single Ln-citrate adsorption mode on the $Ag_{11}$ cluster, which does not capture the complexity of the multi-molecular adsorption modes present in experiments(Figure S1). (iii) Inherent limitations of the DFT approach, including influences from the solvent environment and basis set selection, may lead to Raman shift errors that can be scaled using frequency scaling factors[19,56,57]. In addition, the effect of the ECP-size on the simulation results was verified to be at acceptable levels (Figures S1 and S2). Thus, although the absolute intensities from the DFT-calculated SERS spectra may not be directly comparable to the experimental values, the scaled Raman shifts provide a reliable basis for accurate peak assignment.

We also analyzed the effect of the excitation wavelength on the SERS spectrum of each individual Ln-citrate complex. The relative intensity of the peaks at approximately 1065 and 1315 $cm^{-1}$ under 488 nm excitation was similar to that under 532 nm excitation. However, the relative intensity of the peaks at approximately 1315 and 1485 $cm^{-1}$ differed significantly at these two excitation wavelengths. This is consistent with our previous study[19], which indicated that as the excitation wavelength moves further away from the LSPR peak (around 400 nm), the peak around 1315 $cm^{-1}$ becomes stronger relative to the peak near 1485 $cm^{-1}$. Furthermore, at an excitation wavelength of 660 nm, the signal-to-noise ratio decreased, leading to a reduction in spectral quality while the relationship of relative intensities of each peaks became even more pronounced. The characteristic peak around 1485 $cm^{-1}$ shifted to 1491 $cm^{-1}$, and its relative intensity is lower than that of the peak near 690 $cm^{-1}$. This significantly lower SERS peak intensity under 660 nm excitation compared with those under 488 and 532 nm excitation can be explained by two reasons. The first reason is that the 660 nm laser has a lower frequency, resulting in a lower Raman scattering intensity than those at excitation wavelengths of 488 nm and 532 nm. The Raman scattering intensity $I_{mn}$, corresponding to the transition from state m to state n, is described by the following equations[58,59]:

$$I_{mn} = \frac{128\pi^5}{9C^4}(v_i \pm v_{mn})^4 I_i \sum_{p\sigma} \left|(\alpha_{\rho\sigma})_{mn}\right|^2 \tag{2}$$

$$(\alpha_{\rho\sigma})_{mn} = \frac{1}{h}\sum_e \left[\frac{\langle m|\mu_\sigma|e\rangle\langle e|\mu_\rho|n\rangle}{v_{em}-v_i+\mathrm{i}\Gamma_e} + \frac{\langle m|\mu_\rho|e\rangle\langle e|\mu_\sigma|n\rangle}{v_{en}+v_i+\mathrm{i}\Gamma_e}\right] \tag{3}$$

Where, $I_i$ is the intensity of the incident light with frequency $v_i$. $v_{mn}$represents the Raman shift. $\alpha_{\rho\sigma}$ denotes the component $\rho\sigma$ of the Raman scattering tensor. The summation e includes all quantum mechanical eigenstates of the molecule. $v_{em}$ and $v_{en}$ are the transition frequencies from state m to e and from state e to n, respectively. The terms $\langle m|\mu_\sigma|e\rangle$, $\langle e|\mu_\rho|n\rangle$, $\langle m|\mu_\rho|e\rangle$ and $\langle e|\mu_\sigma|n\rangle$ represent the components of the transition electric dipole moment. $\mu_\sigma$ and $\mu_\rho$ are the electric dipole moment operators in the $\sigma$ (Raman excitation) and $\rho$ (Raman scattering) direction, respectively. $\Gamma_e$ denotes the damping constant of the state. Equation (2) indicates that the Raman scattering intensity is proportional to the fourth power of the scattered light frequency $v_i \pm v_{mn}$. As $v_i \gg v_{mn}$, the Raman scattering intensity is considered proportional to the fourth power of the incident light frequency $v_i$.The second reason is that under 660 nm excitation, the LSPR intensity in the range corresponding to the Raman shift is lower compared to that under 488 nm and 532 nm excitation, resulting in a lower SERS intensity. Notably, only the SERS spectrum of Ce-citrate exhibited a broad peak near 527 cm$^{-1}$ under 660 nm excitation, although the mechanism responsible for its appearance remains unclear.

**Ion-dependent variation of SERS spectra**

Furthermore, we compared the relative intensities of the peaks around 1065, 1315 and 1485 $cm^{-1}$ in the experimental SERS spectra of various Ln-citrate complexes at different excitation wavelengths (Figure 4). The relative intensities of the SERS peaks, SERS intensity ratios, were obtained by normalizing the peak intensity around 1065 or 1485 $cm^{-1}$ to the peak near 1315 $cm^{-1}$. Notably, due to limitations in our experimental conditions, SERS detection of $Pm^{3+}$ was not performed. However, to facilitate analysis of the effect of the number of unpaired 4f electrons of $Ln^{3+}$ on the SERS spectra, $Pm^{3+}$ was included as a blank reference. Figures 4a-4c show the values of the SERS intensity ratio $I_{1065}/I_{1315}$ under 488, 532, and 660 nm excitation, respectively. $I_{1065}/I_{1315}$ did not vary with the excitation wavelengths. Conversely, except for an anomaly with $Ce^{3+}$, which will be discussed later, $I_{1065}/I_{1315}$ decreased with increasing number of unpaired 4f electrons, particularly when that number exceeded 4. Thus, $I_{1065}/I_{1315}$ contains information on the type of $Ln^{3+}$ ion. However, $I_{1065}/I_{1315}$ is nearly identical for $Ln^{3+}$ ions with similar electronic configurations, limiting its use to distinguishing only ions with significantly different electronic configurations, i.e. $La^{3+}$ and $Gd^{3+}$[19].

Figures 4d-4f show the values of the SERS intensity ratio $I_{1485}/I_{1315}$ under 488, 532, and 660 nm laser excitation, respectively. $I_{1485}/I_{1315}$ increased as the excitation wavelength decreased. The trends $I_{1065}/I_{1315}$ and $I_{1485}/I_{1315}$ are consistent with those observations in our previous work. As mentioned earlier, the SERS peaks around 1065, 1315 and 1485 $cm^{-1}$ were assigned as $\gamma(CH_2)+\nu(C–O\cdots Ln)$, $\nu_{sym}(COO^{-})+\gamma(CH_2)$, and $\nu_{asym}(COO^{-})+\gamma(CH_2)$, respectively. According to the CT mechanistic model[16,17], all three SERS peaks reflect excitation energy dependency. $I_{1065}/I_{1315}$ and $I_{1485}/I_{1315}$ illustrate the relationship between peak intensities at approximately 1065 and 1485 $cm^{-1}$ and that around 1315 $cm^{-1}$ as the excitation wavelength changes. Notably, only the peak around 1485 $cm^{-1}$ exhibited a significant response to wavelength variation, with its intensity increasing as the wavelength decreased. This trend was consistently observed for the SERS spectra of all Ln-citrate complexes; thus, it can be considered to originate from the intrinsic vibrational characteristics of citrate. As citrate does not exhibit absorption near 488 nm[60] and $\nu_{asym}(COO^{-})$ includes the asymmetric vibrational mode of $COO^{-}$ coordinated to Ag[19,56], the wavelength-dependent variation of $I_{1485}/I_{1315}$ can be attributed to the additional enhancement of $\nu_{asym}(COO^{-}) + \gamma(CH_2)$ due to the CT effect. In addition, at all excitations, $I_{1485}/I_{1315}$ generally decreased as the number of unpaired 4f electrons in $Ln^{3+}$ ions increased. However, this trend did not hold for Ce-citrate under 532 and 660 nm excitation, and Gd-citrate under 488, 532, and 660 nm excitation.

At all excitation wavelengths, Ce-citrate exhibited the highest $I_{1065}/I_{1315}$, surpassing that of La-citrate. However, under 488 nm excitation, this ratio did not show a significant increase compared with those under 532 and 660 nm excitations. Although statistical variations may exist because of the limited number of spectra, at least six measurements confirmed that the $I_{1065}/I_{1315}$ of Ce-citrate is wavelength-dependent. This suggests that Ce-citrate exhibits an additional response to non-resonant wavelengths. However, due to experimental constraints, the underlying mechanism remains unclear and will be explored in future studies.

**Analysis of the possible mechanism underlying the ion-dependent variation of SERS**

To explore the possible mechanism underlying the ion-dependent variation of $I_{1065}/I_{1315}$ and $I_{1485}/I_{1315}$ in the SERS spectra, we further analyzed the influence of the $Ln^{3+}$ ions by comparing the variation with "lanthanide contraction." [1]. Figures 5a and 5b, compare the relationship between lanthanide contraction and $I_{1065}/I_{1315}$ or $I_{1485}/I_{1315}$, respectively, under different excitation wavelengths. Notably, due to the lack of experimental data for $Pm^{3+}$, its probable trend is represented by a dashed line. As shown in Figure 5a, except for $Ce^{3+}$, $I_{1065}/I_{1315}$ decreased with an increase in the number of unpaired 4f electrons in the $Ln^{3+}$ ion. This trend was similar to lanthanide contraction and was not affected by the excitation wavelength. The main vibration in v(C–O···Ln) contributing to the peak around 1065 $cm^{-1}$ originated from the C–O bond. Therefore, except for $Ce^{3+}$, the variation in $I_{1065}/I_{1315}$ induced by the coordination of the C-O bond to different $Ln^{3+}$ ions. This can be explained by two possible mechanisms: (i) For the vibrational mode v(C–O···Ln), whose main vibration comes from v(C–O), when $Ln^{3+}$ is coordinated to O, the overlap between their electron clouds of $Ln^{3+}$ and O increases with the number of unpaired electrons. This causes the negative charges on O to shift toward Ln, reducing the charge difference and electric dipole moment of the C–O bond. The transition electric dipole moment in equation (3) decreases, leading to a lower intensity of the peak near 1065 $cm^{-1}$ relative to that near 1315 $cm^{-1}$. Thus $I_{1065}/I_{1315}$ decreases with an increase in the number of unpaired electrons in $Ln^{3+}$. (ii) Ln coordination causes a significant change in the electron cloud distribution in the C–O···Ln moiety, and the symmetry decreases with an increase in the number of unpaired electrons in $Ln^{3+}$. This further reduces the intensity of the peak near 1065 $cm^{-1}$ relative to that near 1315 $cm^{-1}$.

Under 488 nm excitation, the trend in $I_{1485}/I_{1315}$ was almost consistent with lanthanide contraction, with a slight increase for all ions except $Gd^{3+}$ (Figure 5b). However, under 532 nm excitation, the $I_{1485}/I_{1315}$ values of $Ce^{3+}$ and $Gd^{3+}$ were higher than those of the preceding $Ln^{3+}$ ions. Under 660 nm excitation, the $I_{1485}/I_{1315}$ values of $Ce^{3+}$, $Eu^{3+}$, and $Gd^{3+}$ were also higher than those of the previous $Ln^{3+}$ ions. Therefore, at these two excitation wavelengths, the trend in $I_{1485}/I_{1315}$ differed completely from lanthanide contraction. Two of the three $COO^-$ groups in citrate are coordinated to $Ln^{3+}$; hence, the attraction between $Ln^{3+}$ and the negative charge on oxygen increases with the number of unpaired 4f electrons. This results in a lower dipole moment of the C–O bond and reduces the symmetry of the $COO^-$ electron cloud. Therefore, except for some $Ln^{3+}$ ions, $I_{1485}/I_{1315}$ generally decreases as the number of unpaired electrons increases. However, because the peaks near 1315 and 1485 $cm^{-1}$ represent overlapping SERS signals from the three COO groups of citrate and the peak near 1485 $cm^{-1}$ is additionally enhanced by the CT effect, the influence of $Ln^{3+}$ ions is irregular and showed weak variability.

In addition, although $Ln^{3+}$ ions have significant mass differences, their impact on the vibrational modes of the peaks at approximately 1065,1315 and 1485 $cm^{-1}$ is minimal. Therefore, the mass difference does not significantly affect the SERS spectra.

## Conclusion

In this study, we successfully measured the SERS spectra of selected Ln-citrate complexes (Ln: La, Ce, Pr, Nd, Sm, Eu, and Gd) at a concentration of $1\times10^{-4}$ M under excitation wavelengths of 488, 532, and 660 nm. The experimental spectra showed three major SERS peaks at approximately 1065, 1315, and 1485 $cm^{-1}$, with no detectable resonance Raman effect or fluorescence signal, despite theoretical predictions. The simulated SERS spectra of the Ln-citrate complexes were successfully calculated using an optimized DFT computational method that directly replaced the small-core ECPs in the def2-tzvpd basis set with large-core ECPs, enabling detailed vibrational analysis and peak assignment. SERS intensity ratios were obtained by normalizing the peak intensity I at 1065 or 1485 $cm^{-1}$ to the that near 1315 $cm^{-1}$. $I_{1065}/I_{1315}$ was not affected by the excitation wavelength but depended on the type of $Ln^{3+}$ ion. In contrast, $I_{1485}/I_{1315}$ increased with decreasing excitation wavelength, reflecting additional enhancement by the CT effect. Both $I_{1065}/I_{1315}$ and $I_{1485}/I_{1315}$ decreased as the number of unpaired 4f electrons in the $Ln^{3+}$ ion increased, except for $Ce^{3+}$, $Eu^{3}$, and $Gd^{3+}$. Although this trend was similar to lanthanide contraction, the mechanism was different. Analysis of the $Ln^{3+}$ coordination region suggested that as the number of unpaired 4f electrons increases, $Ln^{3+}$ becomes more strongly attracted to the negative charges on oxygen. Thus, the electric dipole moment of the C–O bond and its symmetry decreases. Because of current experimental limitations, we did not specifically study the unique peak of Ce-citrate or the potential SERRS effect Nd-citrate. These topics will be investigated in our future work. The findings of the present study provide effective guidance and reference for the vibrational spectroscopy studies of other lanthanides and even actinides.

**Acknowledgments**

The authors acknowledge Dr. Tian Lu (Beijing Kein Research Center for Natural Sciences, China) for his fruitful suggestions for the DFT calculations of rare earth elements. The authors acknowledge funding from JSPS KAKENHI Grant-in-Aid for Scientific Research (C), number 21K04935.

**Supporting information**

Effect of ECP-size differences for $Ln^{3+}$ on the calculation results. Effect of adsorption modes on the simulated SERS spectra. SERS peak assignment. Modeling data.

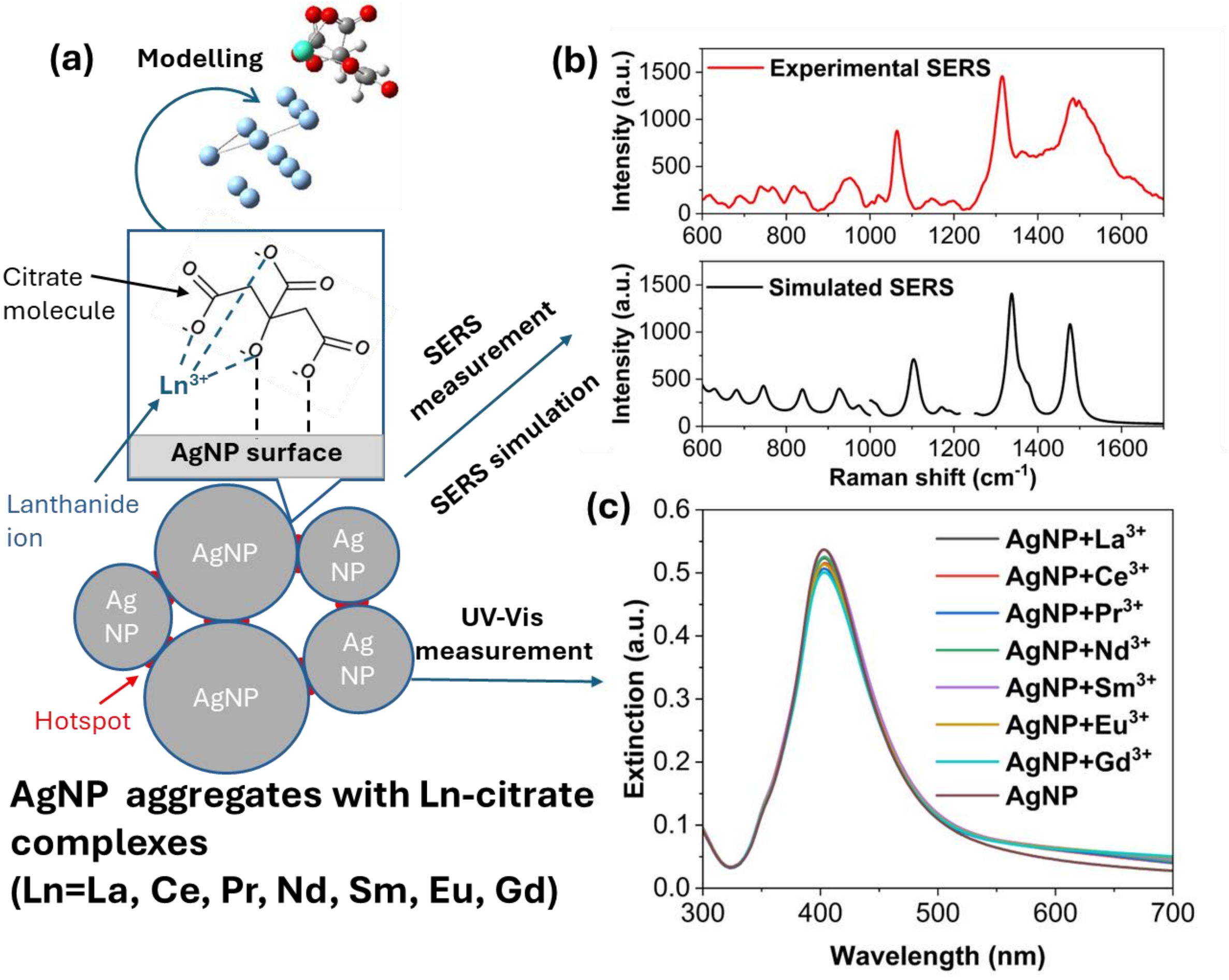

Figure 1. (a) Schematic of the research methods used in this study. (b) Experimental and simulated SERS spectra of Gd-citrate. (c) UV-Vis extinction spectra of original citrate@AgNP and Ln-citrate@AgNP samples.

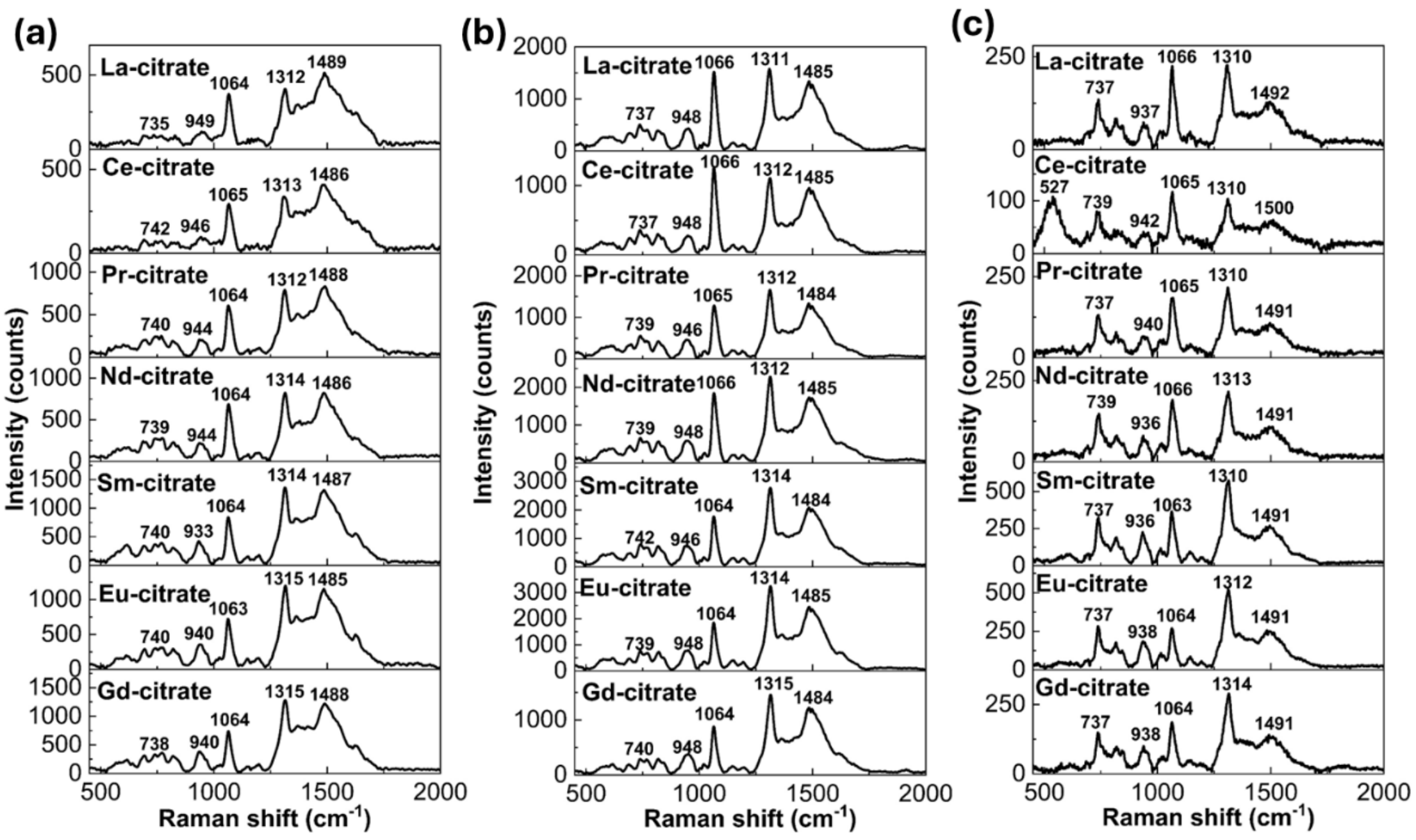

Figure 2. SERS spectra of Ln-citrate complexes under excitation at (a) 488 nm, (b) 532 nm, and (c) 660 nm.

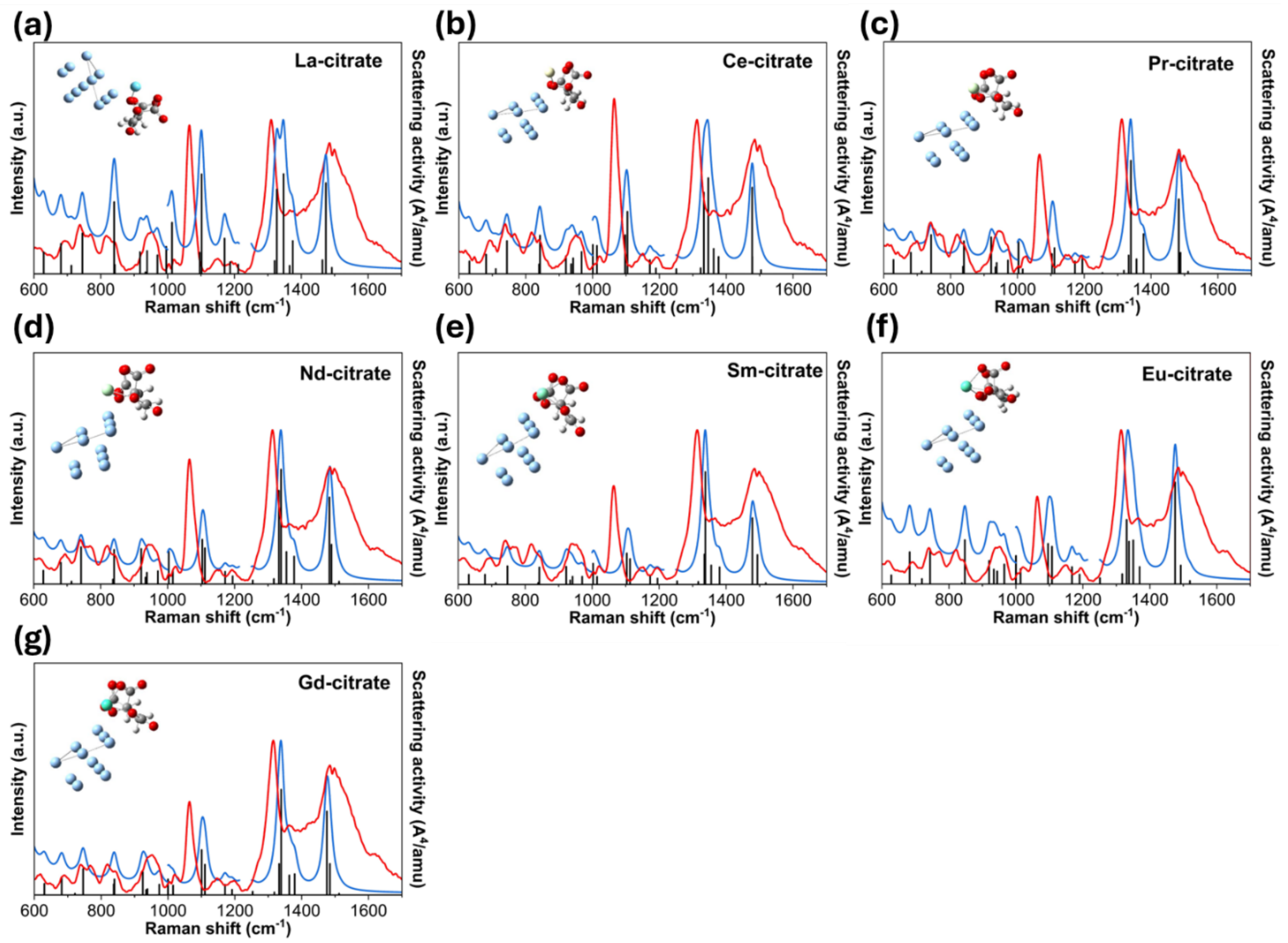

Figure 3. Experimental SERS spectra (red curves) at $1\times10^{-4}$ M, calculated vibrational modes (black curves), and simulated SERS spectra (blue curves) for (a) La-citrate, (b) Ce-citrate, (c) Pr-citrate, (d) Nd-citrate, (e) Sm-citrate, (f) Eu-citrate, and (g) Gd-citrate. As shown in the structures, all simulations were based on a single-molecule SERS model. Some regions between 1000 $cm^{-1}$ and 1213-1250 $cm^{-1}$ were missing due to double scaling when the spectra were exported to Multiwfn. However, this has no effect on the results because there are no vibrational modes in these missing regions.

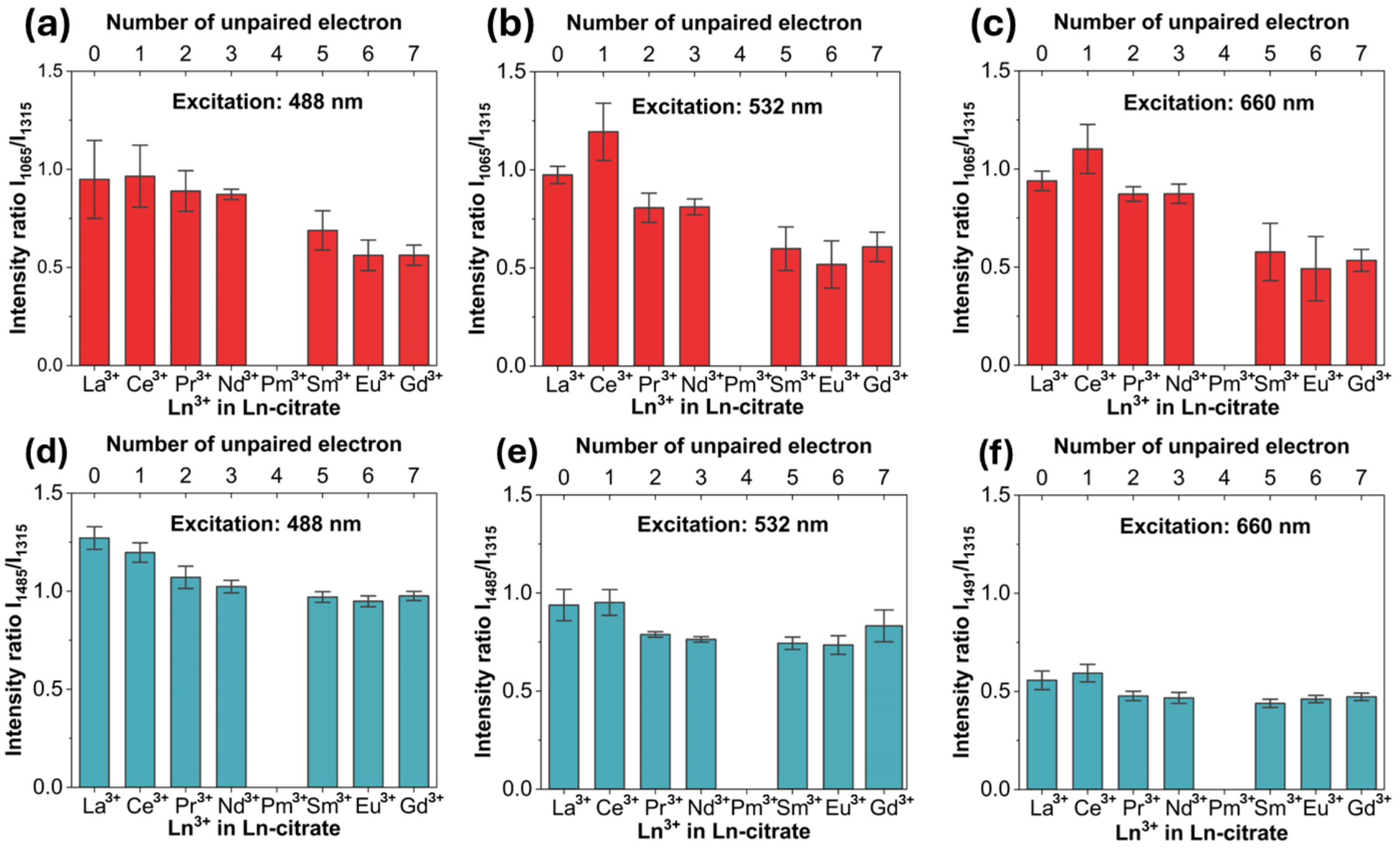

Figure 4. (a-c): SERS intensity ratio $I_{1065}/I_{1315}$ under excitation at (a) 488 nm, (b) 532 nm, and (c) 660 nm. (d-f): Normalized intensity of $I_{1485}/I_{1315}$ under laser excitation at (d) 488 nm, (e) 532 nm, and (f) 660 nm. Notably, when the excitation wavelength is 660 nm, the peak around 1485 cm$^{-1}$ shifted to 1491 cm$^{-1}$. The SERS intensity ratios were obtained by normalizing the peak intensity around 1065 or 1485 cm$^{-1}$ to that near 1315 cm$^{-1}$. The measurement times (N) for La–citrate through Gd–citrate were N = 9, 8, 6, 6, 6, 6, and 7, respectively, under 488 nm excitation. N = 7, 9, 6, 6, 6, 6, and 8, respectively, under 532 nm excitation; and N = 7, 8, 6, 6, 6, 6, and 6, respectively, under 660 nm excitation. Error bars indicate ± SD

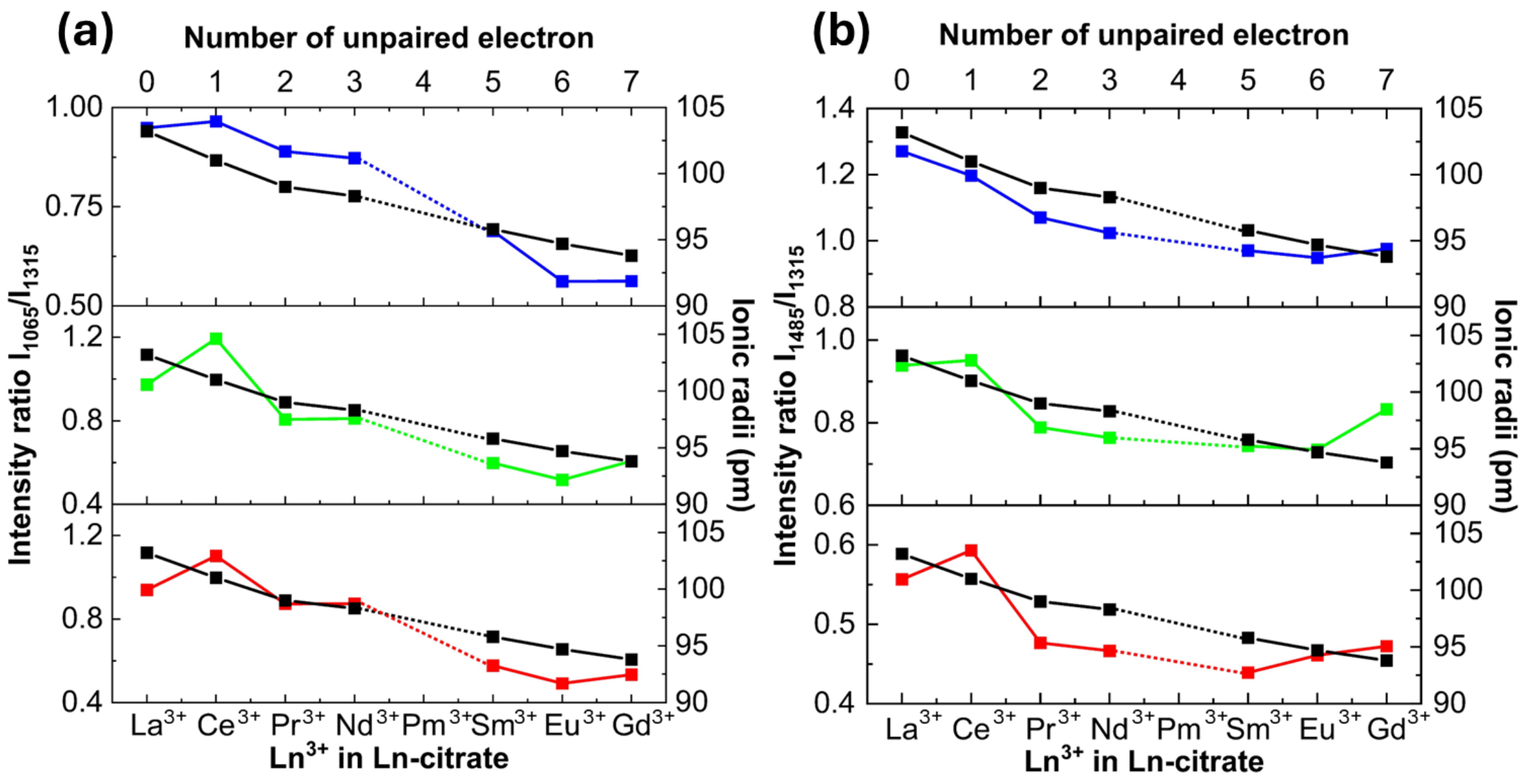

Figure 5. Comparison of lanthanide contraction (black line) with the SERS intensity ratios: (a) $I_{1065}/I_{1315}$ and (b) $I_{1485}/I_{1315}$, measured under 488 nm (blue line), 532 nm (green line), and 660 nm (red line) excitation.

Table 1. Basic information on $Ln^{3+}$[1,21,52]

| Type of $Ln^{3+}$ | Electronic configuration | Ionic radius (pm) | Total spin S | Absorption wavelength (nm) | Emission wavelength (nm) |
|---|---|---|---|---|---|
| $La^{3+}$ | $[Xe]4f^0$ | 103.2 | 0 | None | None |
| $Ce^{3+}$ | $[Xe]4f^1$ | 101 | 1/2 | 215, 240, 257, 273 | 320, 339 |
| $Pr^{3+}$ | $[Xe]4f^2$ | 99 | 1 | 442, 590 | 482, 607, 614, 641 |
| $Nd^{3+}$ | $[Xe]4f^3$ | 98.3 | 3/2 | 355, 455, 525, 580, 680 | 864, 895, 1053, 1330 |
| $Sm^{3+}$ | $[Xe]4f^5$ | 95.8 | 5/2 | 343, 405, 522 | 560, 596, 606, 644, 653, 702 |
| $Eu^{3+}$ | $[Xe]4f^6$ | 94.7 | 3 | 316, 396 | 573, 580, 588, 616, 630, 640, 690, 700 |
| $Gd^{3+}$ | $[Xe]4f^7$ | 93.8 | 7/2 | 275, 308 | 305, 312 |

Table 2. Peak assignments of Ln-citrate complexes

| Type of Ln-citrate | Raman shift ($cm^{-1}$) | | | | | | |
|---|---|---|---|---|---|---|---|
| | 620, 700 | 740 | 820 | 937, 945 | 1065 | 1315 | 1485(1491*) |
| La-citrate | $\delta(COO^-)$ | $\delta(COO^-)$, $\gamma(COO^-)$ | $\nu(CCCC–O)$ | $\nu(C–COO^-)$, $\delta(CH_2)$ | $\nu(C–O\cdots La)$, $\gamma(CH_2)$ | $\nu_{sym}(COO^-)$, $\delta(CH_2)$ | $\nu_{asym}(COO^-)$, $\gamma(CH_2)$ |
| Ce-citrate | $\delta(COO^-)$ | $\delta(COO^-)$, $\gamma(COO^-)$ | $\nu(CCCC–O)$ | $\nu(C–COO^-)$, $\delta(CH_2)$ | $\nu(C–O\cdots Ce)$, $\gamma(CH_2)$ | $\nu_{sym}(COO^-)$, $\delta(CH_2)$ | $\nu_{asym}(COO^-)$, $\gamma(CH_2)$ |
| Pr-citrate | $\delta(COO^-)$ | $\delta(COO^-)$, $\gamma(COO^-)$ | $\nu(CCCC–O)$ | $\nu(C–COO^-)$, $\delta(CH_2)$ | $\nu(C–O\cdots Pr)$, $\gamma(CH_2)$ | $\nu_{sym}(COO^-)$, $\delta(CH_2)$ | $\nu_{asym}(COO^-)$, $\gamma(CH_2)$ |
| Nd-citrate | $\delta(COO^-)$ | $\delta(COO^-)$, $\gamma(COO^-)$ | $\nu(CCCC–O)$ | $\nu(C–COO^-)$, $\delta(CH_2)$ | $\nu(C–O\cdots Nd)$, $\gamma(CH_2)$ | $\nu_{sym}(COO^-)$, $\delta(CH_2)$ | $\nu_{asym}(COO^-)$, $\gamma(CH_2)$ |
| Sm-citrate | $\delta(COO^-)$ | $\delta(COO^-)$, $\gamma(COO^-)$ | $\nu(CCCC–O)$ | $\nu(C–COO^-)$, $\delta(CH_2)$ | $\nu(C–O\cdots Sm)$, $\gamma(CH_2)$ | $\nu_{sym}(COO^-)$, $\delta(CH_2)$ | $\nu_{asym}(COO^-)$, $\gamma(CH_2)$ |
| Eu-citrate | $\delta(COO^-)$ | $\delta(COO^-)$, $\gamma(COO^-)$ | $\nu(CCCC–O)$ | $\nu(C–COO^-)$, $\delta(CH_2)$ | $\nu(C–O\cdots Eu)$, $\gamma(CH_2)$ | $\nu_{sym}(COO^-)$, $\delta(CH_2)$ | $\nu_{asym}(COO^-)$, $\gamma(CH_2)$ |
| Gd-citrate | $\delta(COO^-)$ | $\delta(COO^-)$, $\gamma(COO^-)$ | $\nu(CCCC–O)$ | $\nu(C–COO^-)$, $\delta(CH_2)$ | $\nu(C–O\cdots Gd)$, $\gamma(CH_2)$ | $\nu_{sym}(COO^-)$, $\delta(CH_2)$ | $\nu_{asym}(COO^-)$, $\gamma(CH_2)$ |

Note: $\nu$ indicates stretching, $\nu_{sym}$ is symmetric stretching, $\nu_{asym}$ is asymmetric stretching, $\delta$ is in-plane bending and rocking, and $\gamma$ is out-of-plane wagging and twisting. The calculated SERS frequencies were scaled by using scaling factors. * Under 660 nm excitation;